\documentclass[iop,revtex4]{emulateapj}   
\usepackage[titletoc,title]{appendix}
\usepackage{lscape}
\usepackage{graphicx,natbib,url,twoopt}
\usepackage[varg]{txfonts}
\usepackage{hyperref}               
\usepackage{blkarray}
\usepackage{kbordermatrix}
\usepackage{bm}
\renewcommand{\kbldelim}{(}
\renewcommand{\kbrdelim}{)}

\bibpunct{(}{)}{;}{a}{}{,}    

\received{receipt date}
\revised{revision date}
\accepted{acceptance date}
\begin{document}  

\title{First assessment of the binary lens OGLE-2015-BLG-0232.}
\author{E. Bachelet$^{R0}$, V. Bozza$^{S0,S1}$, C. Han$^{C0}$, A. Udalski$^{O0}$, I.A. Bond$^{M0}$, J.-P. Beaulieu$^{P0,P1}$, R.A. Street$^{R0}$, J.-I Kim$^{C1}$\\ 
and,\\
D. M. Bramich$^{R1}$,
A. Cassan$^{R2}$,
M. Dominik$^{R3,\dagger}$,
R. Figuera Jaimes$^{R7,R3,R4}$,
K. Horne$^{R3}$,
M. Hundertmark$^{R5,R6}$,
S. Mao$^{R7,R8,R9}$
J. Menzies$^{R10}$,
C. Ranc$^{R2}$,
R. Schmidt$^{R6}$,
C. Snodgrass$^{R11}$,
I. A. Steele$^{R12}$,
Y. Tsapras$^{R6}$,
J. Wambsganss$^{R6}$,
\\
(The RoboNet collaboration)\\
P. Mr{\'o}z$^{O0}$,
I. Soszy{\'n}ski$^{O0}$,
M.K. Szyma{\'n}ski$^{O0}$,
J. Skowron$^{O0}$,
P. Pietrukowicz$^{O0}$,
S. Koz{\l}owski$^{O0}$,
R. Poleski$^{O1}$,
K. Ulaczyk$^{O0}$,
M. Pawlak$^{O0}$,
\\
(The OGLE collaboration)\\
F. Abe$^{M2}$
R. Barry$^{M3}$
D. P.~Bennett$^{M3,M10}$
A. Bhattacharya$^{M3,M10}$
M. Donachie$^{M4}$
A. Fukui$^{M5,M13}$
Y. Hirao$^{M1}$
Y. Itow$^{M2}$
K. Kawasaki$^{M1}$
I. Kondo$^{M1}$
N. Koshimoto$^{M11,M12}$
M. Cheung Alex Li$^{M4}$
Y. Matsubara$^{M2}$
Y. Muraki$^{M2}$
S. Miyazaki$^{M1}$
M. Nagakane$^{M1}$
N. J. Rattenbury$^{M4}$
H. Suematsu$^{M1}$
D. J. Sullivan$^{M6}$
T. Sumi$^{M1}$
D. Suzuki$^{M9}$
P. J. Tristram$^{M7}$
A. Yonehara$^{M8}$
\\
(The MOA collaboration)\\
}

\affil{$^{R0}$Las Cumbres Observatory, 6740 Cortona Drive, Suite 102, Goleta, CA 93117 USA}
\affil{$^{R1}$New York University Abu Dhabi, PO Box 129188, Saadiyat Island, Abu Dhabi, UAE}
\affil{$^{R2}$Sorbonne Universit\'es, UPMC Univ Paris 6 et CNRS, UMR 7095, Institut d'Astrophysique de Paris, 98 bis bd Arago, 75014 Paris, France}
\affil{$^{R3}$SUPA, School of Physics \& Astronomy, University of St Andrews, North Haugh, St Andrews KY16 9SS, UK}
\affil{$^{R4}$European Southern Observatory, Karl-Schwarzschild-Str. 2, 85748 Garching bei M\"unchen, Germany}
\affil{$^{R5}$Niels Bohr Institute \& Centre for Star and Planet Formation, University of Copenhagen, {\O}ster Voldgade 5, 1350 - Copenhagen K, Denmark}
\affil{$^{R6}$Zentrum f{\"u}r Astronomie der Universit{\"a}t Heidelberg, Astronomisches Rechen-Institut, M{\"o}nchhofstr. 12-14, 69120 Heidelberg, Germany} 
\affil{$^{R7}$Pysics Departement and Tsinghua Centre for Astrophysics, Tsinghua University, Beijing 100084, China}
\affil{$^{R8}$National Astronomical Observatories, Chinese Academy of Sciences, 20A Datun Road, Chaoyang District, Beijing 100012, China}
\affil{$^{R8}$Jodrell Bank Centre for Astrophysics, School of Physics and Astronomy, The University of Manchester, Oxford Road, Manchester M13 9PL, UK}
\affil{$^{R9}$South African Astronomical Observatory, PO Box 9, Observatory 7935, South Africa}
\affil{$^{R10}$Planetary and Space Sciences, Department of Physical Sciences, The Open University, Milton Keynes, MK7 6AA, UK}
\affil{$^{R11}$Astrophysics Research Institute, Liverpool John Moores University, Liverpool CH41 1LD, UK}

\affil{$^{C0}$Department of Physics, Chungbuk National University, Cheongju 28644, Korea}
\affil{$^{C1}$Korea Astronomy and Space Science Institute, Daejon 34055, Korea}

\affil{$^{O0}$Warsaw University Observatory, Al.~Ujazdowskie~4, 00-478~Warszawa,Poland}
\affil{$^{O1}$Department of Astronomy, Ohio State University, 140 W. 18th Ave.,Columbus, OH  43210, USA}

\affil{$^{M0}$Institute of Natural and Mathematical Sciences, Massey University, Auckland 0745, New Zealand}
\affil{$^{M1}$Department of Earth and Space Science, Graduate School of Science, Osaka University, Toyonaka, Osaka 560-0043, Japan}
\affil{$^{M2}$Institute for Space-Earth Environmental Research, Nagoya University, Nagoya 464-8601, Japan}
\affil{$^{M3}$Code 667, NASA Goddard Space Flight Center, Greenbelt, MD 20771, USA}
\affil{$^{M4}$Department of Physics, University of Auckland, Private Bag 92019, Auckland, New Zealand}
\affil{$^{M5}$Subaru Telescope Okayama Branch Office, National Astronomical Observatory of Japan, NINS,
3037-5 Honjo, Kamogata, Asakuchi, Okayama 719-0232, Japan}
\affil{$^{M6}$School of Chemical and Physical Sciences, Victoria University, Wellington, New Zealand}
\affil{$^{M7}$University of Canterbury Mt.\ John Observatory, P.O. Box 56, Lake Tekapo 8770, New Zealand}
\affil{$^{M8}$Department of Physics, Faculty of Science, Kyoto Sangyo University, 603-8555 Kyoto, Japan}
\affil{$^{M9}$Institute of Space and Astronautical Science, Japan Aerospace Exploration Agency, 3-1-1 Yoshinodai, Chuo, Sagamihara, Kanagawa, 252-5210, Japan}
\affil{$^{M10}$Department of Astronomy, University of Maryland, College Park, MD 20742, USA}
\affil{$^{M11}$Department of Astronomy, Graduate School of Science, The University of Tokyo, 7-3-1 Hongo, Bunkyo-ku, Tokyo 113-0033, Japan}
\affil{$^{M12}$National Astronomical Observatory of Japan, 2-21-1 Osawa, Mitaka, Tokyo 181-8588, Japan}
\affil{$^{M13}$Instituto de Astrof\'isica de Canarias, V\'ia L\'actea s/n, E-38205 La Laguna, Tenerife, Spain}

\affil{$^{S0}$Dipartimento di Fisica "E.R. Caianiello", Universit{\`a} di Salerno, Via Giovanni Paolo II 132, 84084, Fisciano, Italy}
\affil{$^{S1}$Istituto Internazionale per gli Alti Studi Scientifici (IIASS), Via G. Pellegrino 19, 84019 Vietri sul Mare (SA), Italy}

\affil{$^{P0}$ \quad School of Physical Sciences, University of Tasmania, Private Bag 37 Hobart, Tasmania 7001 Australia; jeanphilippe.beaulieu@utas.edu.au}
\affil{$^{P1}$ \quad Sorbonne Universit\' es, UPMC Universit\' e Paris 6 et CNRS, UMR 7095, Institut d'Astrophysique de Paris, 98http://www.plt.axvline/ bis bd Arago, 75014 Paris, France; beaulieu@iap.fr}

\affil{$\dagger$ Royal Society University Research Fellow}

\begin{abstract}
We present an analysis of the microlensing event OGLE-2015-BLG-0232. This event is challenging to characterize for two reasons. First, the light curve is not well sampled during the caustic crossing due to the proximity of the full Moon impacting the photometry quality. Moreover, the source brightness is difficult to estimate because this event is blended with a nearby K dwarf star. We found that the light curve deviations are likely due to a close brown dwarf companion (i.e., $s=0.55$ and $q=0.06$), but the exact nature of the lens is still unknown. We finally discuss the potential of follow-up observations to estimate the lens mass and distance in the future.
\end{abstract}

\section{Introduction}     \label{sec:introduction}
Twenty years after the first exoplanet detection, it is clear that planets are abundant in the Milky Way \citep{Cassan2012,Fressin2013,Bonfils2013,Clanton2016,Suzuki2016}. But the dividing line between 
super-Jupiter and brown dwarfs is still uncertain. \citet{Burrows2001} define brown dwarfs as objects within mass limits $[13,73] ~M_J$. As underlined by \citet{Schlaufman2018},
this definition is problematic because the critical mass for deuterium burning  depends on the object composition \citep{Spiegel2011}. More recently, 
an alternative definition has been proposed based on the formation mechanisms \citep{Schneider2011}: planets are formed by core accretion while brown dwarfs are a result of gas collapse. The former is motivated by exoplanet formation models and by the observational evidence that giants planets tend to form more frequently around metal-rich stars \citep{Mordasini2012,Buchhave2012,Mortier2012}. In contrast, \citet{Latham2002} found no significant correlation between metallicity and stellar binary occurrence. But this definition is also problematic because it is nearly impossible to distinguish the two scenarios observationally \citep{Wright2011, Bryan2018}. Recently, \citet{Schlaufman2018} revisited the mass definition by combining and clustering samples of low-mass stars, brown dwarfs and planets orbiting Solar-type stars and ultimately derived a surprisingly low upper planetary mass limit of $\sim6~M_J$.  

Brown dwarf detections are therefore important to understand the planetary regime boundaries but these objects are intrinsically difficult to detect directly, due to their low-luminosity. Moreover, the radii of brown dwarfs and Jupiter-like planets are very similar due to the degeneracy pressure \citep{Zapolsky1969, Burrows1993}. It is therefore difficult to distinguish them with the transit method alone. Microlensing on the other hand can detect brown dwarfs several {kpc} away, either in binary systems or as single objects \citep{Zhu2016,Chung2017,Schvartzvald2018}, because the method does not need flux measurements from the lens. Several brown dwarfs and brown dwarfs candidates have been discovered through this method \citep{Bachelet2012b,Bozza2012, Ranc2015,Han2016,Poleski2017,Mroz2017}.

In this work, we present the analysis of OGLE-2015-BLG-0232/MOA-2015-BLG-046. The data presented in Section~\ref{sec:obs} show clear signatures of a binary lens event. In Section~\ref{sec:modeling}, we present the modeling procedure and find that the mass ratio of the lens system favors a brown dwarf companion (close model) or a low-mass M dwarf companion (wide model). We present a detailed study of both the microlensing source and the bright blend in Section~\ref{sec:source}. Because no parallax was measured, we discuss in Section~\ref{sec:newobs} the possible follow-up observations to unlock the final solution of this microlensing puzzle.
   
\section{Observations} \label{sec:obs}
The microlensing event OGLE-2015-BLG-0232 ($\alpha=\rm{18^h06^m43.84^s},\delta=\rm{-32^\circ 54^m 27.3^s} ; l=-1^\circ.172199,b=-5^\circ.9060$) was 
an early event of the 2015 microlensing season first discovered by the Optical Gravitational Lens Experiment (OGLE) \citep{Udalski2003b}
on 2015 March 2 UT 17:50 and also detected later by the Microlensing Observations in Astrophysics (MOA) collaboration \citep{Bond2001} as MOA-2015-BLG-046 on 2015 March 10 at UT 16:42.
C. Han first delivered an email alert indicating an ongoing anomaly on 2015 March 15 at UT 02:16. Independently,
the RoboNet team, based on the SIGNALMEN anomaly detector \citep{Dominik2007} and 
the RoboTAP algorithm \citep{Hundertmark2017}, automatically triggered observations on 
the Las Cumbres Observatory network of robotic telescopes \citep{Tsapras2009}. Unfortunately, 
the Moon was nearly full during this period, preventing surveys from acquiring more data 
during the anomaly. This event was also observed in the near infrared by the VISTA Variables in the Via Lactea (VVV) survey \citep{Minniti2010}.
Real-time modeling conducted independently by C.Han and 
V.Bozza indicated that this event was probably due to a low-mass binary lens ($q\sim0.01$).
All teams reprocessed their photometry at the end of the season using the 
difference image analysis (DIA) technique : RoboNet used {\tt DanDIA} \citep{Bramich2008,Bramich2013}, 
OGLE and MOA used their own implementation of DIA \citep{Udalski2015,Bond2001}. The $K$ band of VVV was re-reduced 
using pySIS \citep{Albrow2009}. The VVV pySIS photometry were roughly calibrated to an independent VVV catalog \citep{Beaulieu2016} by adding an offset of 0.6 mag. Note that the $\rm{VVV_{\rm{K}}}$ 
light curve is nearly flat, so we did not use this dataset in the first round of modeling.
In total, 7659 data points are available for the analysis, as summarized in Table~\ref{tab:sumobservations}.
\begin{table*}
  \footnotesize
  
  \centering
  \begin{tabular}{lcccccccc}
    & \\
     \hline\hline
Name&Collaboration&Location&Aperture(m)&Filter&Code&$N_{\rm data}$&Longitude($\deg$)&Latitude($\deg$)\\
    \hline
      & \\
OGLE$_{\rm I}$&OGLE&Chile&1.3&I&Wo\'{z}niak&525&289.307&-29.015\\
MOA$_{\rm Red}$&MOA&New~Zealand&1.8&Red&Bond&6569&170.465&43.987\\
MOA$_{\rm V}$&Boller\&Chivens&New~Zealand&0.6&V&Bond&184&170.465&43.987\\
VVV$_{\rm K}$&VISTA&Chile&4.1&K&pySIS&198&289.6081&-24.616\\
LSCA$_{\rm i}$&RoboNet&Chile&1.0&SDSS-i&{\tt DanDIA}&30&289.195&-30.167\\
LSCB$_{\rm i}$&RoboNet&Chile&1.0&SDSS-i&{\tt DanDIA}&23&289.195&-30.167\\
 LSCC$_{\rm i}$&RoboNet&Chile&1.0&SDSS-i&{\tt DanDIA}&21&289.195&-30.167\\
CPTA$_{\rm i}$&RoboNet&South Africa&1.0&SDSS-i&{\tt DanDIA}&21&220.810&-32.347\\
 CPTB$_{\rm i}$&RoboNet&South Africa&1.0&SDSS-i&{\tt DanDIA}&21&220.810&-32.347\\
 CPTC$_{\rm i}$&RoboNet&South Africa&1.0&SDSS-i&{\tt DanDIA}&12&220.810&-32.347\\
COJA$_{\rm i}$&RoboNet&Australia&1.0&SDSS-i&{\tt DanDIA}&29&149.065&-31.273\\
 COJB$_{\rm i}$&RoboNet&Australia&1.0&SDSS-i&{\tt DanDIA}&18&149.065&-31.273\\
       & \\
    \hline
  \end{tabular}
  \centering
  \caption{Summary of observations.}
  \label{tab:sumobservations}
\end{table*}
\section{Modeling} \label{sec:modeling}
\subsection{Description} \label{sec:modeldescription}
This event is clearly anomalous and real-time models found that a binary lens with a small mass ratio accurately reproduces the observations. A static binary model is described with seven parameters : $t_0$ the time of the minimum impact parameter $u_0$, $t_E=\theta_E/\mu$ the angular Einstein radius crossing time, 
$\rho=\theta_*/\theta_E$ the normalized angular source radius, $s$ the normalized 
projected separation, $q$ the mass ratio between the two lens components and
finally $\alpha$ the lens/source trajectory angle relative to the binary axis. Here, $\mu$ is the relative proper motion
between the source and the lens and $\theta_E$ is the angular Einstein ring, see for example \citet{Gould2000}. Note that we restrict the modeling of the data points to the time window $t\in [2456850,2457200]$ to speed-up the modeling.
For events like OGLE-2015-BLG-0232 that exhibit caustic crossings, the limb-darkening of the source star has to be considered. Unfortunately, in this case, the observations taken around ${\rm HJD \sim 2457087}$ were in ${\rm SDSS-i' }$ band only in order to reduce the impact of the moonlight. Moreover, the caustic crossings are not intensively covered by the data. For these reasons, we investigated a simpler model, the 
Uniform Source Binary Lens (USBL)\citep{Bozza2010, Bozza2012} and use pyLIMA \citep{Bachelet2017} to perform the modeling. 
A detailed description of this binary fitting code is given in Bachelet (2018, in prep).
We did not use the standard grid approach to locate the global minimum, but instead ran a global search on all parameters using the differential evolution method \citep{Storn1997, Bachelet2017}. Briefly, this method uses a set of starting points in parameter space and maintains an ordered population of candidate solutions while exploring potential new solutions by combining existing ones. This algorithm was successfully tested by applying it to previously published events. In practice, we split the parameter space in two regions: $s<1$ and $s>1$. This is motivated by the dramatic change of the caustics topology between these two regimes but also the presence of the close/wide degeneracy, see for example \citet{Erdl1993,Dominik1999,Bozza2000,Cassan2008}.
We ran the algorithm several times and found that it converged to similar solutions.
This event was also modeled in real time by V.Bozza using RTModel\footnote{\url{http://www.fisica.unisa.it/GravitationAstrophysics/RTModel.htm}}. This system uses a different method to explore the parameter space: a template matching approach \citep{Mao1995, Liebig2015}. It also found similar solutions, raising confidence in our results. Results relative to this first exploration can be seen Table~\ref{tab:models}.

\subsection{Error bar rescaling} \label{sec:errorbar}
It is common practice to rescale the uncertainties in microlensing using (in mag unit in the present work):
\begin{equation}
\sigma' = k\sqrt{\sigma^2+e_{\rm{min}}^2}
\label{eq:rescale}
\end{equation}
where $\sigma'$ is the rescaled uncertainty, $k$ and $e_{\rm{min}}$ are parameters that need to be tuned to 
reach a certain metric to optimize. The usual metric used is to force the $\chi^2/\rm{dof}$ for each dataset to 
converge to 1 \citep{Bachelet2012a, Miyake2012, Yee2013}. However, \citet{Andrae2010} show that 
the use of the reduced $\chi^2$, for model diagnostic, is relevant only for linear models, which is not the case in the present work. Instead, they recommend the use of normality tests of residuals, like \citet{Bachelet2015}. 

The physical reasons that motivate the rescaling are to account for photometric low-level systematics and potential underestimation of the uncertainties. There are multiple causes coming from both intrumentation and software reductions. The impact is expected to be different for each dataset and therefore, instead of automatically rescaling the errorbars of each dataset blindly, we assessed wether this was necessary. To do so, we use the approach describe below.
  
First, we rescaled OGLE-IV uncertainties using the custom method of \citet{Skowron2016}\footnote{http://ogle.astrouw.edu.pl/ogle4/errorbars/blg/errcorr-OIV-BLG-I.dat}. We then analyzed the residuals around the best model using three test of normality :  a Kolmogorov-Smirnov test, an Anderson-Darling test  and a Shapiro-Wilk test. We considered to rescale a dataset if any of these test were not successful (i.e, the p-value associated to the test was less than 1\%).
All datasets, except $\rm{MOA_{\rm Red}}$, passed the three normality tests. The majority of datasets present a relative small number of observations ($\le 100$), any deviations to normality would be then hard to detect. On the other hand, it might indicate that uncertainties reproduce the data scatter accurately. Note that the OGLE-IV dataset also passed the three tests after the rescaling process. 

As a secondary check, we follow the same approach as \citet{Dominik2018} and fit the parameters of Equation~\ref{eq:rescale} around the best model from the previous section, using the modified $\chi^{2 \prime} $:

\begin{equation}
 \chi^{2  \prime} = \sum_{i}{{(f_i-m_i)^2}\over{\sigma_i'^2}} + 2\ln( \sigma_i')
\label{eq:pseudochi2}
\end{equation}
with $f_i$ the observed flux, $m_i$ the microlensing model in flux and $\sigma_i'$ is the modified error in flux relative to Equation~\ref{eq:rescale}. It appeared rapidly that the term $e_{\rm{min}}$ was not constrained, due to the relative small range of magnification in the light curves. We therefore delete this term from Equation~\ref{eq:rescale} and fit only the first term $k$. The results presented in the Table~\ref{tab:rescaling} are consistent with the previous analysis and indicate a soft rescaling, with the exception of the $\rm MOA_{\rm Red}$ dataset.

\begin{table}
  \footnotesize
  
  \centering
  \begin{tabular}{lcccccc}
    & \\
     \hline\hline
Name&$N_{\rm data}$&k&$e_{\rm min}$\\
    \hline
      & \\
 OGLE$_{\rm I}^a$&68&$0.97\pm0.06$&0.0\\
 MOA$_{\rm Red}$&467&$3.51\pm0.08$&0.0\\
 MOA$_{\rm V}$&43&$2.1\pm0.2$&0.0\\
 VVV$_{\rm K}$&14&$3.5\pm0.5$&0.0\\
 LSCA$_{\rm i}$&30&$1.1\pm0.1$&0.0\\
 LSCB$_{\rm i}$&23&$1.9\pm0.2$&0.0\\
 LSCC$_{\rm i}$&21&$1.7\pm0.2$&0.0\\
 CPTA$_{\rm i}$&21&$1.1\pm0.1$&0.0\\
 CPTB$_{\rm i}$&21&$1.2\pm0.1$&0.0\\
 CPTC$_{\rm i}$&12&$1.2\pm0.2$&0.0\\
 COJA$_{\rm i}$&29&$1.7\pm0.1$&0.0\\
 COJB$_{\rm i}$&18&$2.1\pm0.3$&0.0\\
       & \\
    \hline
  \end{tabular}
  \centering
  \caption{Error bar rescaling coefficients used in this paper. $^a$ Note that the $\rm{OGLE_I}$ uncertainties receive a special treatment before this rescaling step, see text.}
  \label{tab:rescaling}
\end{table}

\subsection{Results} \label{sec:results}
\begin{figure*}    
  \centering
  \includegraphics[width=18cm]{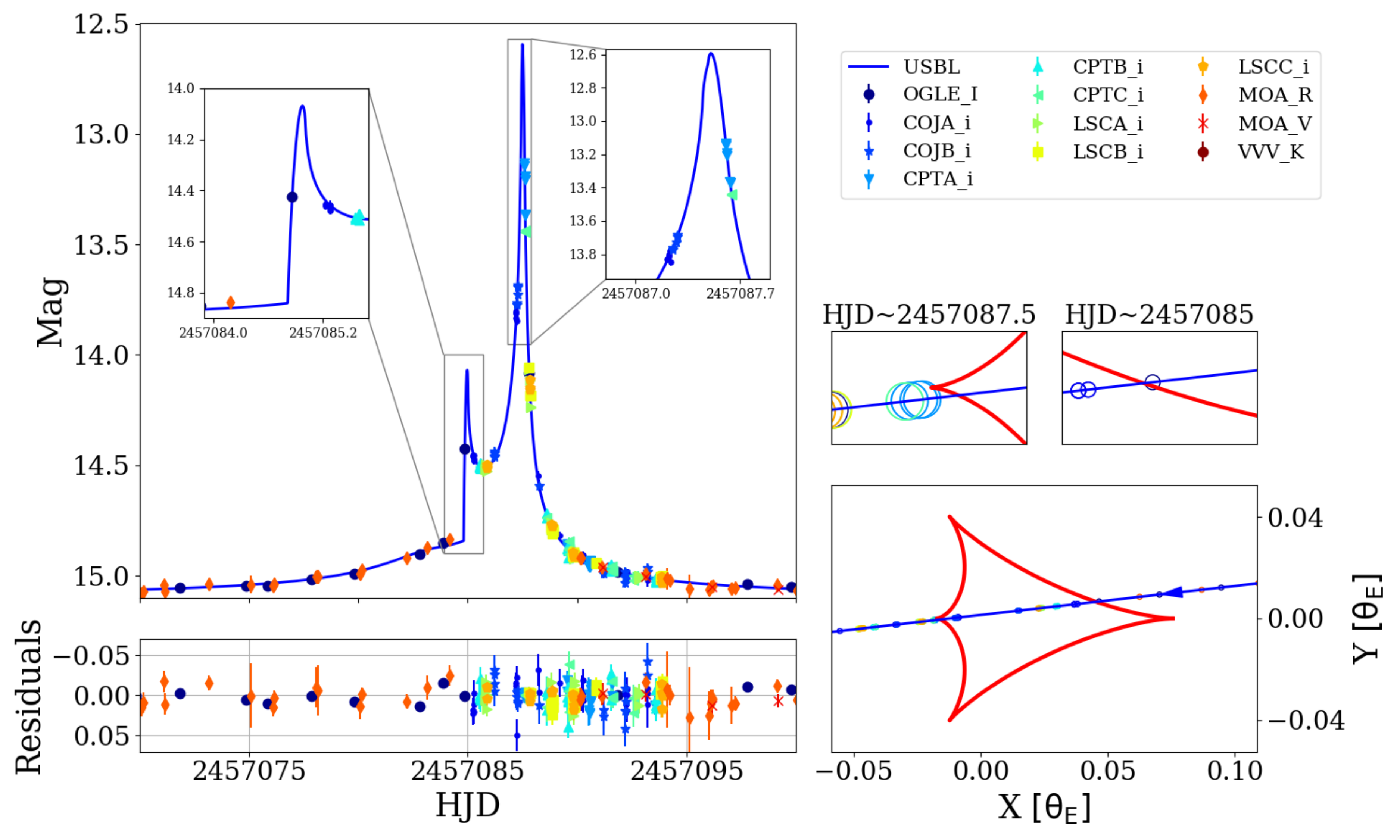}
    \caption{\textit{Left}: Light curves and best-fit model for OGLE-2015-BLG-0232. \textit{Right}: Central caustic (red curve), source trajectory (blue line), and source positions at the epoch of observations. The insets show zooms around the caustic crossings. There was one ${\rm OGLE_I}$ measurement during the caustic entry and four ${\rm CPTA_i}$ points (only two are visible in the inset) during the cusp exit. This allows a reasonable constraint on the normalised source radius $\rho$.}
    \label{fig:lightcurve}
\end{figure*}

Both algorithms 
converged to models with similar geometries:  the strong anomalies seen in Figure~\ref{fig:lightcurve} are due to a central caustic crossing. However the data do not constrain strongly the models, leading to significant difference in the model parameters given in the Table~\ref{tab:models}. To obtain a more comprehensive picture, we run two sets of Monte-Carlo Markov Chain (MCMC) explorations around these best models, using the {\bf emcee} algorithm \citep{Foreman2013} implemented in pyLIMA. Note that during this optimization process, we modified the model parameters so that we model $t_c$ and $u_c$, the time and closest approach to the central caustic, instead of $t_0$ and $u_0$. The idea is to use parameters more directly related to the main features of the light curve. This is a standard practice that significantly improves the model convergence \citep{Cassan2008,Han2009a,Penny2014}.

\begin{table*}
  \centering
  \begin{tabular}{l|ccc|ccc}
     
     \hline\hline
Parameters&pyLIMA ($s<1$)&RT Model$^\dag$($s<1$) &MCMC($s<1$) &pyLIMA ($s>1$)&RT Model$^\dag$($s>1$) &  MCMC($s>1$)\\
\hline
&&\\
$t_c-2450000$ & 7087.20(1)&7087.49(4)&7087.2(2)& 7086.68(4)&7086.93(4)&7086.76(8) \\
$u_c$ & -0.00048(4) & 0.00135(7) &-0.0005(6)&0.00320(8) & 0.0020(2)&0.0026(5) \\
$t_E$ & 41.7(3) & 46.1(3) &42(6)&34.7(1) & 35(3)&39(3) \\ 
$\rho~(10^{-4})$ & 9.9(3) &19.9(8) &10(1)& 7.0(5) & 7.0(2)&7.3(9) \\
$s$ & 0.545(2) & 0.699(2) &0.55(7)&3.05(1) & 2.58(2)&2.9(2) \\
$q$ & 0.0597(8) & 0.0180(1) &0.06(2)&  0.338(3) & 0.17(1)&0.24(6)\\
$\alpha$ & -3.031(3) & -3.061(2)&-3.03(2) & 3.017(4) & 3.045(5)&3.008 (9) \\
$\chi^2$ & 766&799& 764&822&843&812\\
&&\\
  \hline 
  \end{tabular}
\\
$^\dag$ The parameters are obtained from the online RTModel website (http://www.fisica.unisa.it/gravitationAstrophysics/RTModel/2015/RTModel.htm).\\
  \centering
  \caption{Close/Wide best models of pyLIMA, RTModel and MCMC explorations. Models from RTModel were used as starting point for a Levenberg-Marquardt (LM) optimization with pyLIMA. Numbers in bracket in the table represents 1 $\protect\sigma$ errors from LM and 68\% range for the MCMC explorations.}
  \label{tab:models}
\end{table*}

The geometry of the best fitting model is sensitive to the close/wide degeneracy \citep{Griest1998,Bozza2000,Dominik2009}. However, close models are slightly favored. The mass ratio of this event is not well constrained. This is due to a lack of observations during the anomaly, especially during the central caustic entrance and exit. 

We tried to model second-order effects, such as annual parallax and the orbital motion of the lens \citep{Gould1992,Dominik1998,Albrow2000,Gould2004,Bachelet2012a}. Due to the relatively short timescale of the event and the 
relatively low coverage of the anomaly features, these second order effects were not constrained well enough to be considered as a solid detection.


\section{Properties of the source} \label{sec:source}

\subsection{Optical observations} \label{sec:optical}
\begin{figure*}    
  \centering
  \includegraphics[width=15cm]{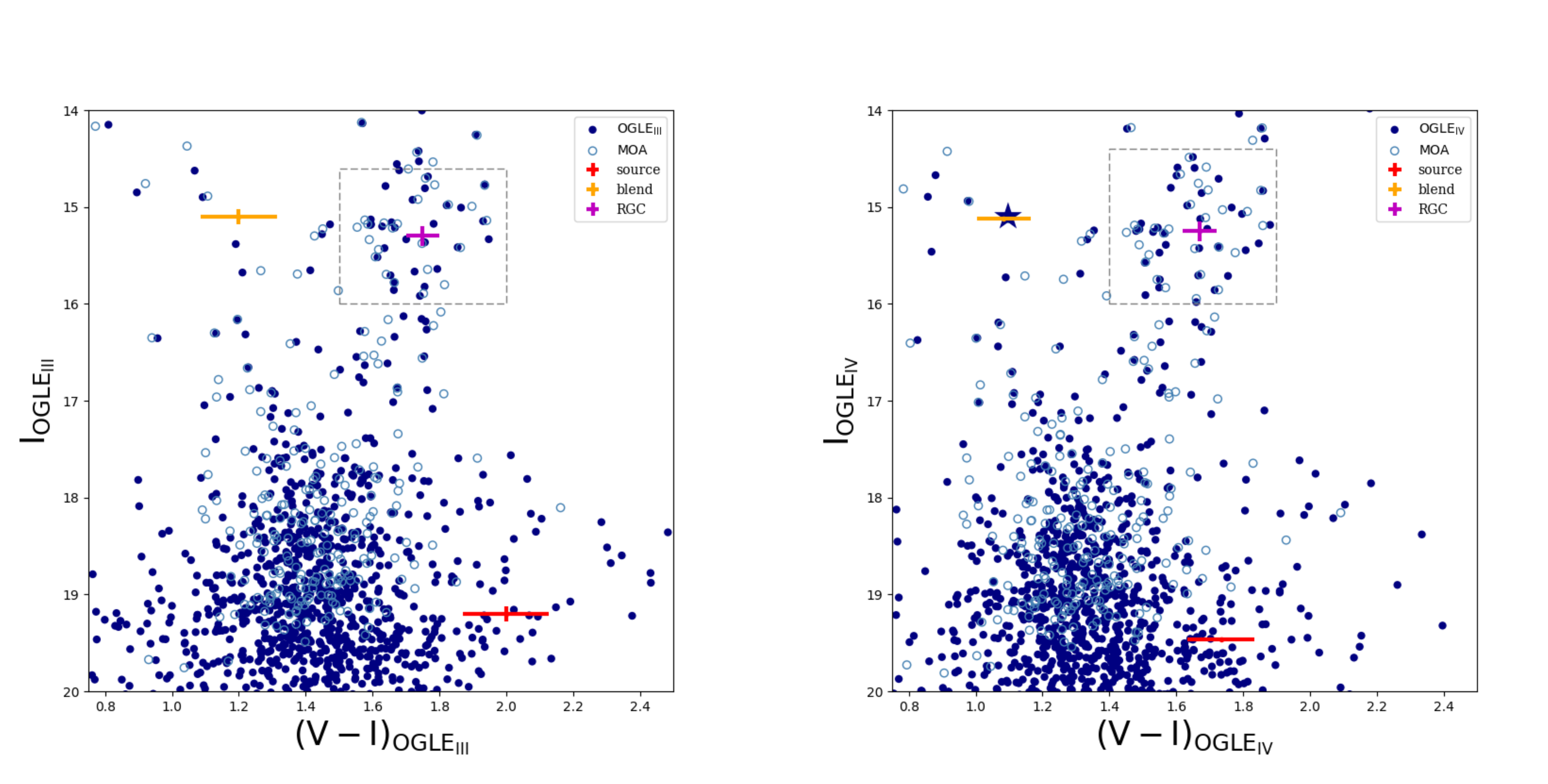}
    \caption{Optical color-magnitude diagrams of stars within 2' of the line of sight of this event. OGLE and the transformed MOA are in blue (filled and empty respectively), the source is in red, the blend is in orange, and the position of the RGC is in magenta. The star symbol represents the star presented in the Table~\ref{tab:astrometry}. The grey squares represent the region used to estimate the position of the RGC. Left :$\rm{OGLE_{III}}$ photometric system (i.e, Johnson-Cousins \citep{Szymanski2011}). Right: instrumental $\rm{OGLE_{IV}}$ photometric system.  }
    \label{fig:CMDVI}
\end{figure*}

Following \citet{Bond2017}, we calibrated the $\rm{MOA}_{\rm{R}}$ and $\rm{MOA}_{\rm{V}}$ magnitudes 
to the $\rm{OGLE}_{\rm{III}}$ system using the relation in the Appendix. The resulting color-magnitude diagram (CMD thereafter) is presented in Figure~\ref{fig:CMDVI}, and we summarize information from the various catalogs used in Table~\ref{tab:astrometry}. We found that the color of the red giant clump (RGC) centroid is $(V-I)_{\rm{RGC}} = 1.75 \pm 0.05$ mag and its brightness is $I_{\rm{RGC}} = 15.3\pm 0.1$ mag. Knowing the intrinsic color of the RGC $(V-I)_{\rm 0,RGC}=1.06$ mag and its 
intrinsic brightness $I_{\rm 0,RGC}=14.45$ mag \citep{Nataf2012}, we estimate the absorption $A_I = 0.9 \pm 0.1$ mag and the extinction $E(V-I) = 0.69 \pm 0.05$ mag toward the microlensing event. We found a good agreement with an independent determination 
using the Interstellar Extinction Calculator on the OGLE website\footnote{http://ogle.astrouw.edu.pl/}, based 
on \citet{Nataf2012} and \citet{Gonzalez2012}, with $A_I = 0.79 \pm 0.1$ mag and $E(V-I)=0.68 \pm 0.05$ mag. From the best model and the color transformations in the Appendix, the source magnitudes are $V_{\rm{s,OGLE_{\rm{III}}}} = 21.2 \pm 0.1$ mag and $I_{\rm{s,OGLE_{\rm{III}}}} = 19.15\pm0.09$ mag (and a color of $(V-I)_{\rm{s,OGLE_{\rm{III}}}}= 2.0 \pm 0.1$ mag). In principle, it is possible 
to obtain a model-independent color using linear regression between two bands $\lambda_1$ and $\lambda_2$ since the microlensing magnification is achromatic \citep{Dong2006,Bond2017}:
\begin{equation}
f_{\lambda_1} = {{f_{s,\lambda_1}}\over{f_{s,\lambda_2}}} (f_{\lambda_2}-f_{b,\lambda_2})+f_{b,\lambda_1}
\end{equation}
where $f_s$ and $f_b$ are the source and blending flux respectively. However, this requires simultaneous observations which are difficult in practice. Here, we consider $\rm{MOA}_{\rm{R}}$ and $\rm{MOA}_{\rm{V}}$ as simultaneous if the acquisition time was within 15 minutes. We found a model independent source color of $(V-I)_{{\rm{s,{\rm{OGLE_{\rm{III}}}}}}}=2.0 \pm 0.1$ mag, in agreement with the previous estimation. Finally, we obtained the intrisic color $(V-I)_{{\rm{o,s,{\rm{OGLE_{\rm{III}}}}}}}= 1.4 \pm 0.1$ mag and brightness $I_{\rm{o,s,OGLE_{\rm{III}}}} = 18.4\pm 0.1$ of the source in the OGLE-III system (i.e., in the Johnson-Cousins system).

Because this event was also observed by OGLE-IV, we conducted a similar analysis using the OGLE-IV CMD. The corresponding CMD is presented in Figure~\ref{fig:CMDVI}. In this CMD, we found that the color of the red giant clump (RGC) centroid is $(V-I)_{\rm{RGC}} = 1.67 \pm 0.05$ mag and its brightness is $I_{\rm{RGC}} = 15.3\pm0.1$ mag. The best model and the $V$-band transformation in Equation~\ref{eq:OGLEIV} lead to  $V_{\rm{s,OGLE_{\rm{IV}}}} = 21.2 \pm 0.1$ mag and $I_{\rm{s,OGLE_{\rm{IV}}}} = 19.47\pm0.01$ mag (and a color of $(V-I)_{\rm{s,OGLE_{\rm{IV}}}}= 1.7 \pm 0.1$ mag). Assuming the source suffers the same extinction as the RGC, we measured an offset between the source and the RGC $\Delta((V-I)_{\rm{s,OGLE_{\rm{IV}}}},I_{\rm{s,OGLE_{\rm{IV}}}})=(0.03\pm0.1,4.2\pm0.1)$. However, the OGLE-IV system is not perfectly calibrated, and the difference in the colors need to be multiplied by a factor $0.93$ (for the CCD 24 of the OGLE camera mosaic) \citep{Udalski2015}. Based on the OGLE-IV CMD, the source color is $(V-I)_{{\rm{o,s,{\rm{OGLE_{\rm{IV}}}}}}}= 1.09 \pm 0.1$ mag and the brightness is $I_{\rm{o,s,OGLE_{\rm{IV}}}} = 18.7\pm 0.1$ mag.

While the two studies converge to a similar conclusion, we use for the source properties $(V-I)_{{\rm{o,s,{\rm{OGLE_{\rm{IV}}}}}}}= 1.09 \pm 0.1$ mag
and $I_{\rm{o,s,OGLE_{\rm{IV}}}} = 18.7\pm 0.1$ mag, because they rely on a single color transformation and also because the color term in Equation~\ref{eq:OGLEIV} is smaller than the one in Equation~\ref{eq:OGLEIII}. From optical observations, the source is probably a K-dwarf \citep{Bessell1988} or, potentially, a K subgiant that lies behind the Galactic Bulge.

Using \citet{Kervella2008} and the optical color, we can obtain the angular source radius $\theta_*$. We obtain 13\% precision on $\theta_*=0.8~\pm 0.1 ~\rm{\mu as}$. Finally, we can then estimate the angular Einstein ring radius $\theta_E=\theta_*/\rho=0.8\pm0.2$ mas (using the best model) and $\mu=7.0 \pm 3$ mas/yr. This provides one mass and distance constraint to the lens system since \citep{Gould2000}:
\begin{equation}
M_{tot} = {{\theta_E^2}\over{\kappa\pi_{rel}}}
\label{eq:mldl}
\end{equation}
with $\pi_{rel} = {{1-x}\over{D_s x}}$au, $x=D_l/D_s$ (the distance to the lens and the source respectively) and the constant $\kappa  = 8.144 ~ {\rm mas.M_\odot^{-1}}$.
\subsection{Near infrared} \label{sec:NIR}
\begin{figure}    
  \centering
  \includegraphics[width=9cm]{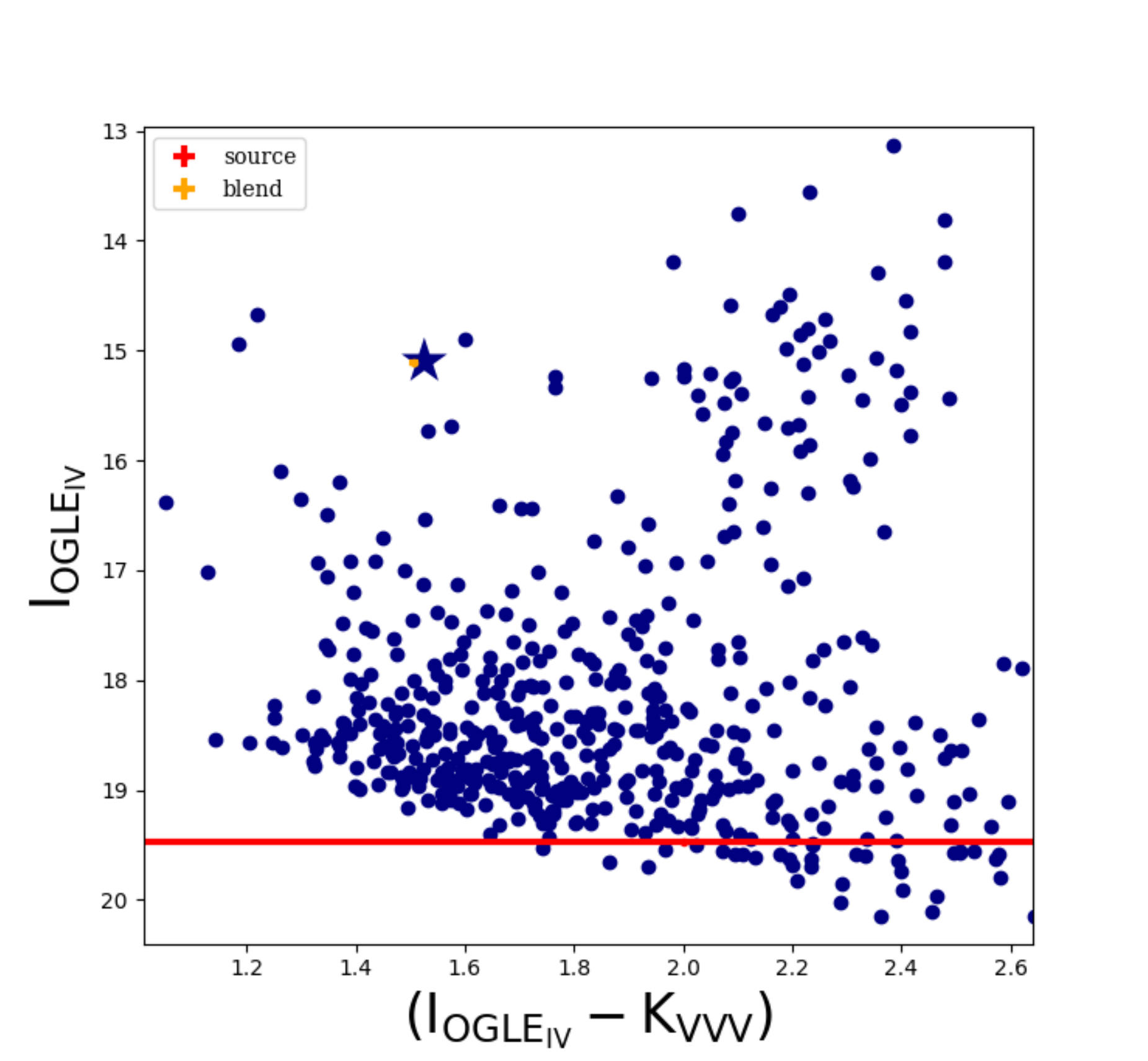}
    \caption{Color magnitude diagram of stars within 2' of the line of sight to this event, using $\rm{I_{{\rm{{\rm{OGLE_{\rm{IV}}}}}}}}$ and $\rm{K_{\rm{VVV}}}$. }
    \label{fig:VVVCMD}
\end{figure}

Thanks to VVV observations, we can perform a similar study using $K$-band data and construct a near-infrared CMD, as shown in Figure~\ref{fig:VVVCMD}.
\citet{Gonzalez2012} provide extinction maps toward the Galactic Bulge. Using their online tool \footnote{http://mill.astro.puc.cl/BEAM/calculator.php},
we found $A_K=0.10\pm0.06$ mag and $E(J-K)=0.19\pm0.11$ mag. This agrees relatively well with the 3D Maps toward  
the Galactic Bulge of \citet{Schultheis2014} (i.e., $E(J-K)=0.30\pm0.06$ mag and $A_K=0.16\pm0.04$ mag assuming 
\citet{Nishiyama2009} extinction law). From the best model, the source brightness is $K_{s,s} = 17 \pm 1$ mag and the blend brightness is $K_{s,b} = 13.61 \pm 0.03$ mag. The relatively low precision on the source magnitude in $K_s$ is again due to the lack of observations during the event high-magnification pahse of the event.  Unfortunately, the maximum observed magnification was only $A\sim 1.6$, while the secondary maximum observed magnification was $A\sim1.05$. The color of the source is $\rm{(I_{{\rm{{\rm{OGLE_{\rm{IV}}}}}}}-K_{\rm{VVV}})}= 2 \pm 1$ mag, leading to an extinction corrected color of $(I_{{\rm{{\rm{OGLE_{\rm{IV}}}}}}}-K_{\rm{VVV}})_o=2.0-A_I+A_K = 1 \pm 1$ mag, and a magnitude of $K_{0,\rm{VVV}}=17\pm 1$ mag. Using \citet{Bessell1988}, we found that this color is consistent with the optical colour and corresponds to a K-type source star. 
\subsection{Does the source belong to the Sagittarius Dwarf Galaxy?} \label{sec:discussion}
\begin{figure}    
  \centering
  \includegraphics[width=9cm]{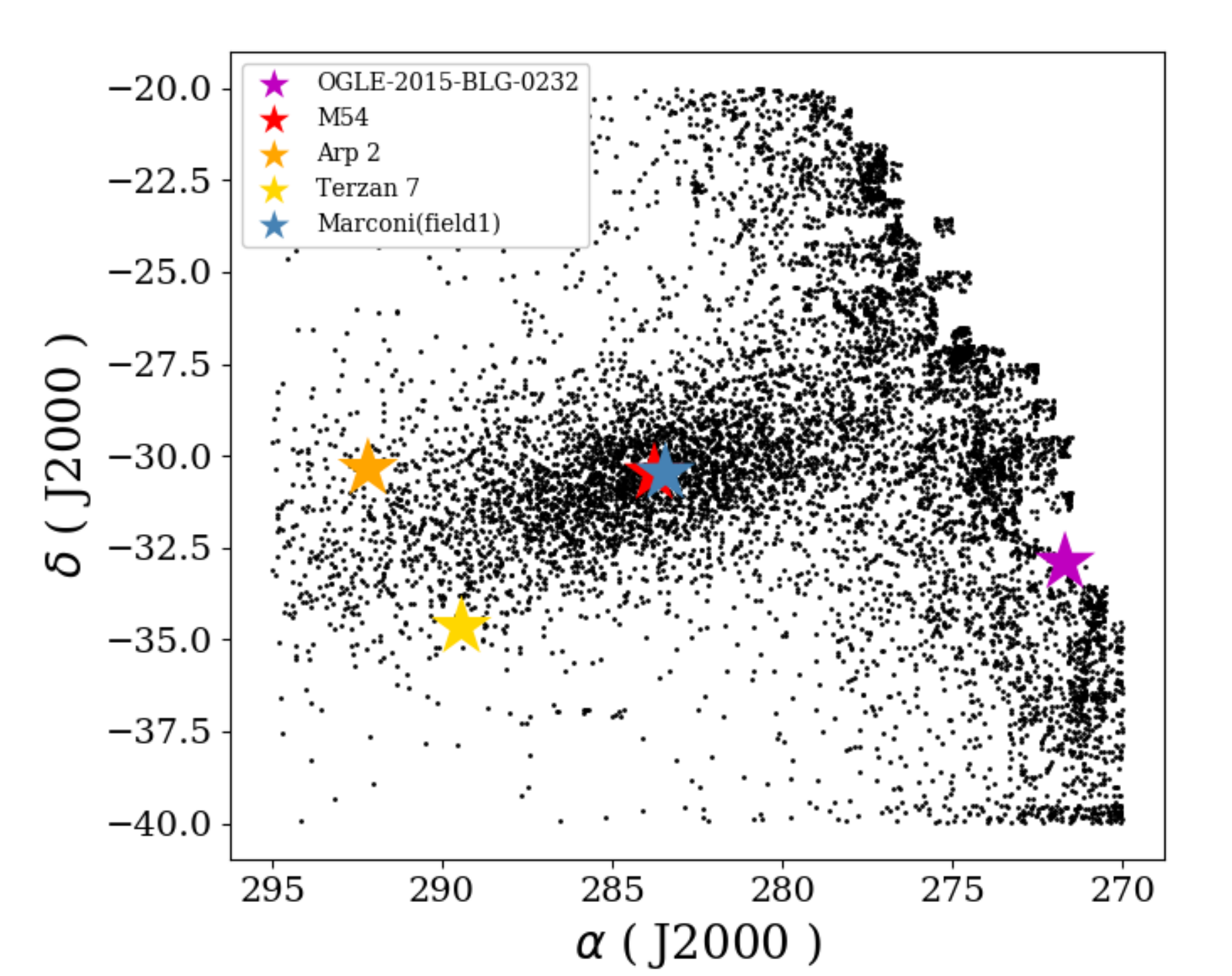}
    \caption{Map of the Sagittarius Dwarf galaxy from the 2MASS catalog \citep{Cutri2003, Skrutskie2006}.}
    \label{fig:sgr}
\end{figure}

Due to the relatively large galactic latitude of the event (i.e, $b=-5^\circ.9060$), the line of sight does not go through much of the Galactic Disk. This raises the possibility that the source is located in the stream of the Sagittarius Dwarf galaxy \citep{Ibata1994}. If this were the case, the source would be located very far away, $D_s\sim25$ kpc. \citet{Cseresnjes2001} predicted that events due to the Sagittarius dwarf galaxy should represent roughly 1\% of the total events detected toward the Galactic Bulge fields each year. They also predicted that these events should mainly occur for main-sequence source stars with $V\ge 21$ mag and that the median Einstein ring radius crossing time would be 1.3 times larger than the one observed for Milky Way sources. To test this, we constructed a map of the Sagittarius Dwarf galaxy in Figure~\ref{fig:sgr}. We followed the method of \citet{Majewski2003} and selected stars with $E(B-V)<0.555$, $0.95<(J-K_s)_o<1.10$ and $10.5<K_{s,o}<12$ combined with the extinction maps from \citet{Schlegel1998} (with a low resolution of 0.5 deg)\footnote{We use the python  implementation available at https://github.com/gregreen/dustmaps}. However, the line of sight ($\ell=-1^\circ.17,b=-5^\circ.90$) is quite distant from the highest density of the Sagittarius Dwarf galaxy:  M54. The Sagittarius dwarf star population has been studied in great detail, see for example \citet{Marconi1998,Monaco2002,Monaco2004,Guiffrida2010}. Several features can be used to distinguish stars from the Milky Way and the dwarf galaxy. In particular, the CMD of the dwarf galaxy presents several horizontal branches and red-giant branches, signatures of different star populations. The optical CMD of OGLE-2015-BLG-0232 does not show these signatures, indicating that there is no significant contamination from the dwarf galaxy. 

Due to the large distance to the center of the Sagittarius Dwarf galaxy ($\ge 10^\circ$) and the absence of particular features in the CMD, we discount this hypothesis and assume that the source star belongs to the Milky Way.
\section{Information on the blend} \label{sec:simbad}
Results from our modeling indicate that this event was highly blended. It is clear from Figure~\ref{fig:CMDVI} and Figure~\ref{fig:VVVCMD} that the blend belongs to the foreground stars branch of the CMD, indicating a close blend. In the following, we consider the blend as a single star and neglect the potential contamination from the source because the blend ratio is substantial with $g\sim50$.
\subsection{Gaia measurements} \label{sec:Gaia}
The Gaia mission \citep{GAIA2016,GAIA2018,Luri2018} recently released a vast catalog of parallax and proper motions measurements for more than a billion of stars. In addition to this goldmine, effective temperatures, radii and luminosities are also estimated. We summarized the Gaia measurements for OGLE-2015-BLG-0232 in Table~\ref{tab:astrometry}. Recent studies indicate biases in Gaia parallax measurements of several $\mu$as \citep{Lindegren2018,Zinn2018,Riess2018}. We therefore use the estimation of the blend distance $D_b=1023_{-75}^{+86}$ pc by \citet{BailerJones2018}, and so the blend is a late-type G or an early-type K dwarf.  For this target, we also found $T_{\rm eff} = 4707_{-228}^{+269}$ K, $R=1.0_{-0.1}^{+0.1}$ $R_\odot$, $\mathcal{L}=0.42_{-0.07}^{+0.07} \mathcal{L}_\odot$ and ultimately estimated the mass of this blend $M\sim \mathcal{L}^{1/4}\sim 0.8 M_\odot$, typical of a K-dwarf. However, \citet{Andrae2018} note that these parameters are estimated by neglecting the extinction toward the target. While this approximation is reasonable for this target because the blend is relatively close and the extinction along the line of sight is relatively small, these fundamental parameters are probably biased.

The brightnesses of the blend in the Gaia bands are  $G = 15.918 \pm 0.001$ mag, $G_{BP} = 16.48 \pm 0.01$ mag and $G_{RP} = 15.15 \pm 0.01$ mag. Using the system transformation in the Appendix, we convert these magnitudes to the Johnson-Cousins system to find $V = 16.25 \pm 0.05$ mag and $I = 15.08\pm0.05$ mag. Given the blend distance, we assumed half extinction and found an intrinsic color $(V-I)_{o,b,G}= 1.33-0.69/2 = 1.0 \pm 0.1$ and brightness $I_{o,b,G}= 15.08-0.79/2 = 14.7 \pm 0.1$ mag, typical of a K2 dwarf star \citep{Bessell1988}. Using the color-effective temperature relation of \citet{Casagrande2010}, the blend effective temperature is $T_{\rm eff}=4900\pm400$ K. We estimated the blend physical radius $R_b = 0.8 \pm 0.1$ $R_\odot$, the luminosity $\mathcal{L}=0.3 \pm 0.1 \mathcal{L}_\odot$ and finally derived the blend mass $M_b\sim 0.7 M_\odot$ \citep{Boyajian2012}. Knowing that the angular radius of the blend is $\theta_b= 4.8 \pm 0.5 \mu as$ \citep{Kervella2008}, one can derive an indepedent estimate of the blend distance $D_b=800 \pm 200$ pc, in good agreement with the Gaia parallax measurement.

If the blend were the lens and assuming that the source is at 8 kpc, a blend mass of $M_b\sim 0.7 M_\odot$ at a distance $D_b\sim1000$ pc, the angular Einstein ring would be $\theta_{E,b}\sim2.2$ mas. This is in strong disagreement with the value of $\theta_E=0.8 \pm 0.2$ mas derived in Section~\ref{sec:optical}. This is a first clue that the bright blend is likely not the lens.

From Table~\ref{tab:astrometry}, the proper motion of the bright blend is $\bm{\mu_G}(N,E) = (2.2 \pm 0.1, 9.9 \pm 0.1)$ mas/yr. The speed of the Sun in the Galactic frame is $\bm{V_\odot}(U,V,W) \sim (11,12,7) + (0,220,0)$ km/s \citep{Fich1989,Schonrich2010}: the first term is the intrinsic Sun velocity and the second term is the speed of the Galactic disk in the Galactic coordinates system. Assuming the source is at 8 kpc, the expected proper motion of the source is about $\bm{\mu_s}(l,b) \sim (-6 \pm 3, 0 \pm 3)$ mas/yr, see \citet{Kuijken2002} and \citet{Kozlowski2006} for the estimation of the uncertainties. The Galactic proper motion transform to $\bm{\mu_s}(\alpha,\delta) \sim (-3 \pm 3, -5 \pm 3)$ mas /yr \citep{Binney1998,Poleski2013,Bachelet2018}. Therefore, if the bright blend were the lens, one would expect a relative proper motion of $\bm{\mu_{rel}}(N,E) = \mu_s-\mu_G \sim (-5 \pm 3, -15 \pm 3)$ mas/yr. The relative proper motion would be ${\mu_{rel}}=16 \pm 4$ mas/yr, in disagreement with the estimation $\mu_{rel} = 7\pm 3$ mas/yr of the Section~\ref{sec:optical}. This is the second clue that the blend is not the lens.
\subsection{Blend brightness from models} \label{sec:blendmodel}
Using our best-fit model and the color relationships given in the Appendix, we derived the brightnesses of the blend: $\rm{I}_{\rm{b,OGLE_{\rm{IV}}}} = 15.1163\pm0.0007$ mag, $\rm{V}_{\rm{b,OGLE_{\rm{IV}}}} = 16.23\pm0.08$ mag and  $\rm{K_{b,\rm{VVV}}}=13.61\pm 0.03$ mag.  Assuming that the blend suffers half the extinction, we found that the blend brightness is  $I_{o,b,{\rm{{\rm{OGLE_{\rm{IV}}}}}}}=14.7 \pm 0.1$ mag and the blend color is $(I_{{\rm{{\rm{OGLE_{\rm{IV}}}}}}}-K_{\rm{VVV}})_{o,b} = 1.1 \pm 0.1$ mag, consistent with its being an early K-dwarf \citep{Bessell1988}. This is in good agreement with the Gaia measurements. 
\subsection{Astrometry} \label{sec:astrometry}
Toward the Galactic Bulge and for stellar masses, microlensing occurs when the alignement between the lens and the source is less than a few mas. We therefore compared the position of centroids between the baseline object and the magnified source from the OGLE-IV images. In pixel coordinates, the magnified source has an offset of $\Delta(N,E) = (78\pm 45,78\pm 35)$ mas from the bright blend centroid (the precision of the bright blend centroid is about 0.05 pixel, i.e. $\sigma (N,E) = (13,13)$ mas). The two positions are different enough (i.e., $1.5 \sigma$) to assume that this is the third clue that the blend is likely not the lens.

\begin{table*}[h]
  \centering
  \begin{tabular}{lclllccc}
    & \\
     \hline\hline
     
Catalog&Source ID&Epoch&RA(J2000)&DEC(J2000)&Parallax&$\mu_{\alpha}$&$\mu_{\delta}$ \\
&&&$^\circ$&$^\circ$&mas&mas/yr&mas/yr\\    
     \hline

Gaia&4042761215742767360&J2015.5&271.68268633(1)&-32.90760309(1)&0.96(7)&9.9(1)&2.0(1)\\
MOA &965&-&271.68263(4) & -32.90764(3)& -&-&- \\
OGLE-III &90793&J2002.46&271.68267(4) &-32.90761(3)& -&-&- \\
OGLE-IV (baseline) &58780&J2011.4&271.68267(4)&-32.90758(3)& -&-&- \\
OGLE-IV (source) &-&J2011.4&271.68269(4)&-32.90720(3)& -&-&- \\
VVV &2508&J2010&271.68262&-32.90763& -&-&- \\
PPMXL &4938889137283654706&J1991.21&271.68268(2)&-32.90760(2)& -&10.6(5.2)&6.3(5.2) \\
    \hline
  \end{tabular}
  \centering
  \caption{Astrometry of the target in MOA, OGLE-III, OGLE-IV, VVV and Gaia catalogs. OGLE-III catalog of the field is from \citet{Szymanski2011}. OGLE-IV catalog is available online \protect\footnote{\url{http://ogle.astrouw.edu.pl/ogle4/ews/2018/ews.html}}.  VVV catalog is from \citet{Beaulieu2016}. The PPMXL catalog is from \citet{Roeser2010}. Numbers in parenthesis are the 1 $\protect\sigma$ uncertainties.}
  \label{tab:astrometry}

\end{table*}

\begin{figure}    
  \centering
  \includegraphics[width=9cm]{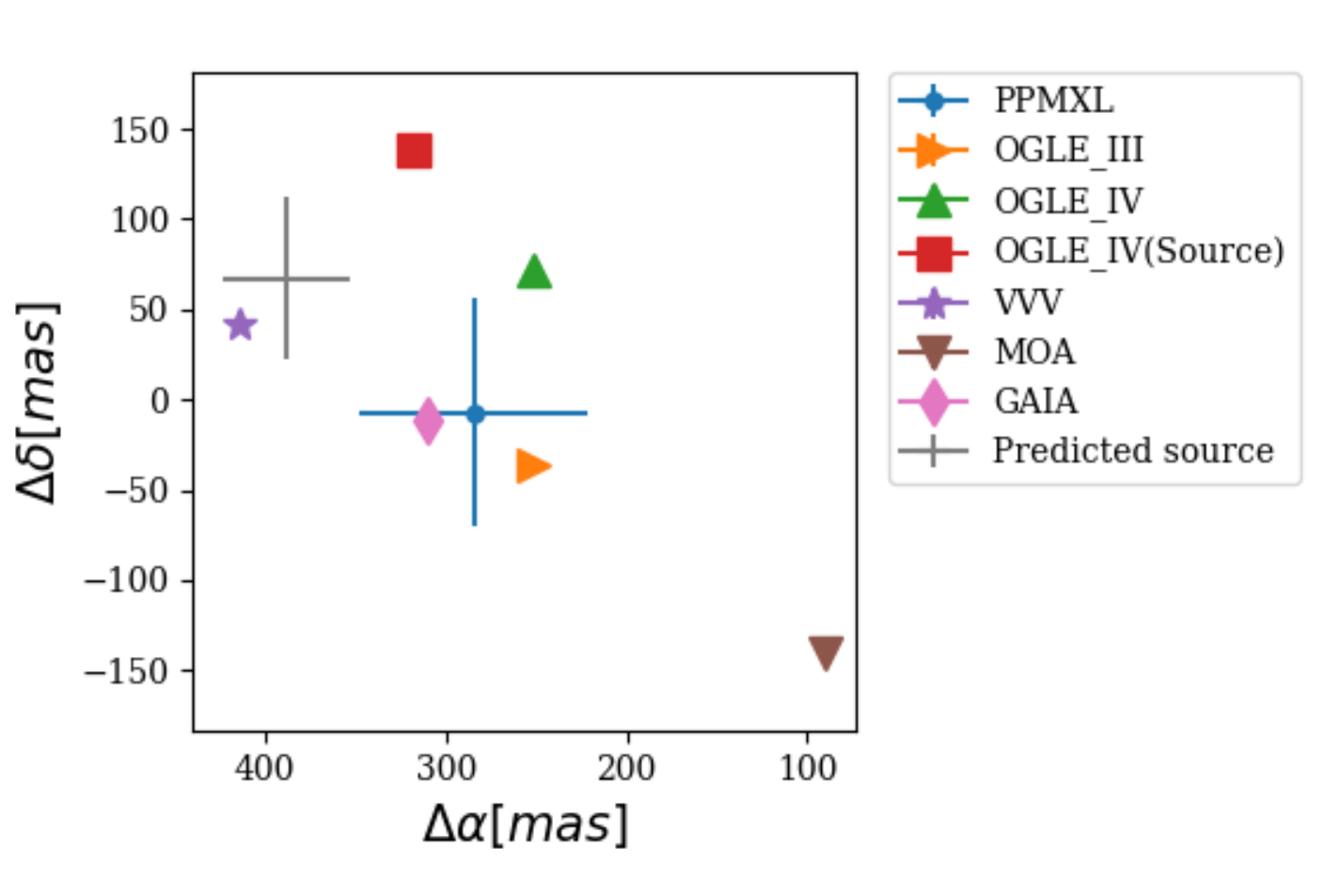}
    \caption{Location of the target from the catalogs listed in the Table~\ref{tab:astrometry}, centered at (RA = $271.6826^\circ$, DEC = $-32.9076^\circ$) (J2000). North is up and East is left. Some uncertainties have been hidden for clarity. The source position measured from OGLE-IV is located at $\Delta(E,N) = (78,78)$ mas (red square) from the OGLE-IV position. The grey cross indicates the prediciton of the source, applying the offset to the Gaia position and assuming $\sigma (N,E) = (45,35)$ mas uncertainties.}
    \label{fig:astrometry}
\end{figure}
\subsection{The lens as a blend companion } \label{sec:blend companion}
In the following, we explore the possibility that the lens is a companion of the blend. From the astrometry offset derived in Section~\ref{sec:astrometry}, we can derive the separation $\delta$  of the 
blend with its potential companion and found $\delta=110$ mas, which corresponds to ${\rm a_{proj}} \sim  110 $ au at 1 kpc. If this potential companion is indeed a component of the lens system, then the mass ratio between the binary blend components is $q_b = (\theta_E/\theta_{E,b})^2 = (0.8/2.2)^2 \sim 0.13$, leading to a potential companion mass of $M_{b,2} = 0.13 \times 0.7 \sim 0.1 M_\odot$. Therefore, such a companion is not bright enough to have been significantly detected. Because the normalised separation between the putative companion and the bright blend is important $s_b=110/2.2 \sim 50$, the hypothetic companion blend could have acted as a binary lens and left no signature of a triple-lens, as observed. 
However, this hypothetic companion would have a similar proper motion as the bright blend and the analysis on the relative proper motions in Section~\ref{sec:Gaia} also apply here. Therefore, the lens as a blend companion hypothesis is unlikely. 
\section{Discussion and potential new clues} \label{sec:newobs}
All available information seem to concur that the blend light is mainly due to a close K dwarf. Both astrometry and the constraint from finite-source effects reject the hypothesis that the bright blend is the lens. The light of the lens is not signifcantly detected and there are no constraints from the microlensing parallax: the distance and exact nature of the lens remains uncertain at the present time. However, considering a large mass range for the lens primary $M_{l,1}\in[0.1,2.0] M_\odot$ (corresponding to $D_l\le 5.5$ kpc according to Equation~\ref{eq:mldl} and $\theta_E=0.8 \pm 0.2$ mas), the companion mass range is ${M_{l,2}\in[6,130] ~ M_{Jup}}$. The lens companion is therefore a massive planet, a brown dwarf or a low-mass M-dwarf if $D_l\le 5.5$ kpc. It the lens is more distant, the primary is probably a stellar remnant, otherwise the lens light would have been detected. This indicates the need for supplementary observations to reveal the nature of the lens OGLE-2015-BLG-0232.

High-resolution imaging is an important tool for microlensing. Several planets have been confirmed using 
space or ground-based facilities and had their measured properties refined, see for example \citep{Batista2015,Bennett2015,Beaulieu2017}.
High-resolution imaging is useful for two reasons. First, it is possible to estimate the source-lens 
proper motion $\mu$ from high-resolution images obtained several years after the microlensing 
event, when the source and the lens are well separated \citep{Batista2015}. High-resolution imaging 
can also provide measurements of source and, sometimes, lens fluxes and therefore tightly constrain the mass-distance relation of the lens 
\citep{Ranc2015,Batista2015,Bennett2015,Beaulieu2017}.

In the case of OGLE-2015-BLG-0232, high-resolution imaging will contribute to confirming/rejecting scenarios and possibly estimate the mass of the lens. The first step will be to challenge the assumption that the blend is a single star. This can be done immediately. Moreover, one can predict a more precise source position based on Gaia astrometry and the measured offset from the OGLE-IV photometry. The predicted position of the source is shown in the Figure~\ref{fig:astrometry}, assuming 26 mas precision on OGLE-IV measurement (i.e., 0.1 pixel). The comparison of the flux at this position in high resolution images with the measured source fluxes from models could place constraints on the nature of the lens.

A second step will be to wait several years for the bright blend leaves the line of sight to obtain more information on the source/lens system. Because $\mu_b=11 \pm 0.2~ {\rm mas/yr}$, the blend is separating faster than the lens/source system $\mu=7.0 \pm 3 ~{\rm mas/yr}$. In a decade, the blend should be about 11 pixels away from the line of sight while the source and the lens separation should be about 7 pixel (for a typical high-resolution pixel scale of 10 mas/pix).

Low-resolution spectroscopy could also confirm the spectral type of the bright blend. Similarly, the study of emission/absorption lines with high-resolution spectroscopy would allow a precise understanding of the blend. Finally, one could combine spectroscopic and photometric information to explore various scenarios in a Bayesian analayis \citep{Santerne2016}. 

\section{Conclusion} \label{sec:conclusion}
We presented an analysis of the binary microlensing event OGLE-2015-BLG-0232. Because the event occurred during full moon, the observations do not constrain much the deviations from the single-lens model. However, results from the modeling favor a close brown dwarf companion (i.e., $s\sim0.55$ and $q\sim0.06$). The source is estimated to be red and faint, probably a K dwarf in the Galactic Bulge. We also tested, and ultimately rejected, the hypothesis that the source belongs to the Sagittarius Dwarf Galaxy. Since the microlensing parallax is not measured, we obtain only one (weak) constraint, from finite-source effects, on the mass and distance of the lens. Based on the recent Gaia DR2 release and OGLE-IV astrometry, we were able to infer that the bright blend is a K dwarf at 1 kpc and is most likely not the lens.
We finally discuss the potential of additional observations to confirm the nature of the blend and ultimately to derive the exact nature of the lens.

\section*{Acknowledgements}
The authors thank the anonymous referee for the constructive comments. This research has made use of NASA's Astrophysics Data System. Work by EB and RAS is support by the NASA grant NNX15AC97G.
Work  by  C.  Han  was  supported  by  the  grant
(2017R1A4A1015178)  of  National  Research  Foundation of Korea. This work makes use of observations from the LCOGT network. The OGLE project has received funding from the National Science Centre,
Poland, grant MAESTRO 2014/14/A/ST9/00121 to AU. The MOA project is supported by JSPS KAKENHI Grant Number JSPS24253004, JSPS26247023, JSPS23340064, JSPS15H00781,  JP16H06287 and JP17H02871. The work by C.R. was supported by an appointment to the NASA Postdoctoral Program at the Goddard Space Flight Center, administered by USRA through a contract with NASA.
This work has made use of data from the European Space Agency (ESA)
mission {\it Gaia} (\url{https://www.cosmos.esa.int/Gaia}), processed by
the {\it Gaia} Data Processing and Analysis Consortium (DPAC,
\url{https://www.cosmos.esa.int/web/Gaia/dpac/consortium}). Funding
for the DPAC has been provided by national institutions, in particular
the institutions participating in the {\it Gaia} Multilateral Agreement.
This publication makes use of data products from the Two Micron All Sky Survey, which is a joint project of the University of Massachusetts and the Infrared Processing and Analysis Center/California Institute of Technology, funded by the National Aeronautics and Space Administration and the National Science Foundation. This research made use of Astropy, a community-developed core Python package for Astronomy (Astropy Collaboration, 2013). This research has made use of the SIMBAD database,
operated at CDS, Strasbourg, France.
The DENIS project has been partly funded by the SCIENCE and the HCM plans of
the European Commission under grants CT920791 and CT940627.
It is supported by INSU, MEN and CNRS in France, by the State of Baden-W\"urttemberg 
in Germany, by DGICYT in Spain, by CNR in Italy, by FFwFBWF in Austria, by FAPESP in Brazil,
by OTKA grants F-4239 and F-013990 in Hungary, and by the ESO C\&EE grant A-04-046.
Jean Claude Renault from IAP was the Project manager.  Observations were  
carried out thanks to the contribution of numerous students and young 
scientists from all involved institutes, under the supervision of  P. Fouqu\'e,  
survey astronomer resident in Chile.  DPB, AB, and CR  were supported by NASA through grant NASA-80NSSC18K0274. 
\appendix

\section{Color transformations} \label{sec:colortransformation}

In this work, we used several color transformations that we summarize here.
First, we calibrated the MOA instrumental magnitudes to the OGLE-III catalog \citep{Udalski2003b,Bond2017} using the relationships :
\begin{equation}
I_{\rm{OGLE_{\rm III}}} = R_{\rm{MOA}} + (27.935 \pm 0.003) + (-0.244 \pm 0.003) (V_{\rm{MOA}} - R_{\rm{MOA}}) \pm 0.08
\end{equation}
\begin{equation}
V_{\rm{OGLE_{\rm III}}} =  V_{\rm{MOA}} + (28.556 \pm 0.002)   + (-0.164 \pm 0.002) (V_{\rm{MOA}} - R_{\rm{MOA}}) \pm 0.08
\label{eq:OGLEIII}
\end{equation}

We also calibrated the MOA instrumental magnitudes to the OGLE-IV system using:
\begin{equation}
I_{\rm{OGLE_{\rm IV}}} = R_{\rm{MOA}} + (27.990 \pm 0.003) + (-0.247 \pm 0.009) (V_{\rm{MOA}} - R_{\rm{MOA}}) \pm 0.08
\end{equation}
\begin{equation}
V_{\rm{OGLE_{\rm IV}}} =  V_{\rm{MOA}} + (28.425 \pm 0.005)   + (-0.062 \pm 0.006) (V_{\rm{MOA}} - R_{\rm{MOA}}) \pm 0.08
\label{eq:OGLEIV}
\end{equation}

We also used the transformation of the 2MASS colors into the the VVV system \citep{Soto2013}:
\begin{equation}
J_{\rm{VVV}} = J_{\rm{2MASS}} -0.077 (J_{\rm{2MASS}} - H_{\rm{2MASS}})
\end{equation}
\begin{equation}
H_{\rm{VVV}} = H_{\rm{2MASS}} + 0.032 (J_{\rm{2MASS}} - H_{\rm{2MASS}})
\end{equation}
\begin{equation}
K_{\rm{VVV}} = K_{\rm{2MASS}} + 0.010 (J_{\rm{2MASS}} - K_{\rm{2MASS}})
\end{equation}

Transformations into the Bessell \& Brett photometric system \citep{Bessell1988} are the revised version \footnote{https://www.ipac.caltech.edu/2mass/releases/allsky/doc/sec6\_4b.html}
of \citet{Carpenter2001}:
\begin{equation}
(K_s)_{\rm{2MASS}} = K_{\rm{BB}} +(-0.039 \pm 0.007) + (0.001 \pm 0.005) (J-K)_{\rm{BB}}
\end{equation}
\begin{equation}
(J-K_s)_{\rm{2MASS}} = (-0.018 \pm 0.007) + (0.001 \pm 0.005) (J)_{\rm{BB}}
\end{equation}

Finally, the transformation of the Gaia DR2 to the Johnson-Cousins system is available online \footnote{https://gea.esac.esa.int/archive/documentation/GDR2/}:

\begin{equation}
G - V_{\rm JC}= -0.01760 -0.006860 (G_{\rm BP}-G_{\rm RP}) -0.1732 (G_{\rm BP}-G_{\rm RP})^2 \pm 0.045858
\end{equation}
\begin{equation}
G - I_{\rm JC}= 0.02085 +0.7419 (G_{\rm BP}-G_{\rm RP}) -0.096311 (G_{\rm BP}-G_{\rm RP})^2 \pm 0.04956
\end{equation}
\bibliographystyle{aa}
\bibliography{biblio_ob150232.bib}
\end{document}